# Physically based alternative to the PE criterion for meteoroids

Manuel Moreno-Ibáñez[1,5], Maria Gritsevich[2,3,4], Josep M. Trigo-Rodríguez[5,6], Elizabeth A. Silber[7,8]

[1]Aistech Space S.L., Rua das Pontes 3, 36350, Nigrán, Galicia, Spain.
manuel.morenoibanez@gmail.com
[2]Finnish Geospatial Research Institute, Geodeetinrinne 2, 02430, Masala, Finland
maria.gritsevich@nls.fi
[3]Department of Physics, University of Helsinki, Gustaf Hällströmin katu 2a, P.O. Box 64, FI-00014 Helsinki, Finland.
[4]Institute of Physics and Technology, Ural Federal University, Mira str. 19. 620002 Ekaterinburg, Russia.
[5]Institute of Space Sciences (ICE, CSIC), Meteorites, Minor Bodies and Planetary Science Group, Campus UAB, Carrer de Can Magrans, s/n E-08193 Cerdanyola del Vallés, Barcelona, Catalonia, Spain.
[6]Institut d'Estudis Espacials de Catalunya (IEEC), Gran Capità 2-4, Ed. Nexus, desp. 201, E-08034 Barcelona, Catalonia, Spain.
[7]AstrumPrime Space Research Inc., Dartmouth, NS, B2W 6C4, Canada
[8]Department of Earth Sciences, Western University, 1151 Richmond St., London, Ontario, N6A 5B7, Canada.

**Abstract**

Meteoroids impacting the Earth atmosphere are commonly classified using the PE criterion. This criterion was introduced to support the identification of the fireball type by empirically linking its orbital origin and composition characteristics. Additionally, it is used as an indicator of the meteoroid tensile strength and its ability to penetrate the atmosphere. However, the level of classification accuracy of the PE criterion depends on the ability to constrain the value of the input data, retrieved from the fireball observation, required to derive the PE value. To overcome these uncertainties and achieve a greater classification detail we propose a new formulation using scaling laws and dimensionless variables that groups all the input variables into two parameters that are directly obtained from the fireball observations. These two parameters, $\alpha$ and $\beta$, represent the drag and the mass loss rates along the luminous part of the trajectory, respectively, and are linked to the shape, strength, ablation efficiency, mineralogical nature of the projectile, and duration of the fireball. Thus, the new formulation relies on a physical basis. This work shows the mathematical equivalence between the PE criterion and the logarithm of $2\alpha\beta$ under the same PE-criterion assumptions. We demonstrate that $\log(2\alpha\beta)$ offers a more general formulation which does not require any preliminary constraint on the meteor flight scenario and discuss the suitability of the new formulation for expanding the classification beyond fully disintegrating fireballs to larger impactors including meteorite-dropping fireballs. The reliability of the new formulation is validated using the Prairie Network meteor observations.

## 1. Introduction

During the past mid-century the observations of fireballs in the Earth atmosphere fostered the development of atmospheric entry models to improve the physical characterization of the impacting bodies. One of the multiple approaches considered that the ability of a meteoroid to penetrate the Earth atmosphere is related to some observable characteristics of its atmospheric flight. A link between these parameters was outlined by Ceplecha & McCrosky (1976), who introduced a compositional and orbital dynamics discriminant called the PE criterion.

This confers the PE criterion with very interesting classification capabilities. On one hand, the PE criterion allows a prompt first order quantification of the meteoroid strength. Since there is clear evidence for different tensile strength properties for meteoroids associated with different streams (Trigo-Rodríguez & Llorca 2006, 2007; Trigo-Rodríguez 2019), the PE criterion has become a common tool (1) to assess a membership of observed meteor events to already identified meteor showers, (2) to suggest the possible existence of new meteor shower streams, and (3) to confirm the existence of meteorite-droppers among meteor shower streams (Brown et al. 2013). On the other hand, the PE criterion can also be combined with other orbital dynamics criteria. For example, it can be used with the Tisserand parameter to gain a more accurate understanding of the cometary or asteroid origin of the event under study (Matlovic et al. 2017).

Although the PE criterion is commonly applied in meteor science, its main drawback is the necessity of an initial assumption, averaging, and rounding of the involved input parameters in order to obtain the remaining parameters in the PE formulation. This eventually leads to a different level of inaccuracies and biases in the results. Despite the PE criterion being extensively used over the years, it has experienced no revision to solve this issue ever since. The scope of this work is to revisit the formulation of this criterion and to introduce a physically based alternative.

The fast-growing computer processing capabilities have fostered the development of alternative fireball modelling strategies. In this regard, to overcome the lack of meteor flight information on relevant body characteristics, Stulov, Mirskii, & Vislyi (1995), Stulov (1997), Gritsevich & Stulov (2006), and Gritsevich (2007) implemented scaling laws and dimensionless parameters to tackle the original Newtonian formulation in a different way. This new formulation, which has been proven and validated in many other works (see Moreno-Ibáñez, Gritsevich & Trigo-Rodríguez (2016) for a review), reduces all the meteor flight modelling unknowns to two parameters, namely $\alpha$ and $\beta$, that have a physical meaning and can be obtained from the meteor observations (Gritsevich 2008a, 2009; Bouquet et al. 2014; Vaubaillon et al. 2015; Trigo-Rodríguez et al. 2015; Sansom et al. 2015; Dmitriev, Lupovka & Gritsevich 2015; Lyytinen & Gritsevich 2016; Meier et al. 2017; Gritsevich et al. 2017). The ballistic coefficient, $\alpha$, is proportional to the mass of the atmospheric column with the cross section $S_e$ along the trajectory divided by the meteoroid's pre-atmospheric mass. In other words, $\alpha$ expresses the drag intensity suffered by the meteor body during its flight. The mass loss parameter, $\beta$, characterizes the mass loss rate of the meteoroid. It can be expressed as the fraction of the kinetic energy per mass unit of the body that is transferred to the body in the form of heat divided by the effective destruction enthalpy.

The suitability to use the parameters $\alpha$ and $\beta$ to classify meteoroid impacts was introduced by Gritsevich, Stulov & Turchak (2009, 2012) and was recently demonstrated on a large observational dataset by Sansom et al. (2019). In particular, using this approach Sansom et al. (2019) identify meteorite-producing fireballs illustrating their 'clustering' on the $\alpha$-$\beta$ plot. It was suggested in Moreno-Ibáñez, Gritsevich & Trigo-Rodríguez (2015, 2017), as a prospect for future research, that the mathematical combination of $\alpha$ and $\beta$ could also lead to the revision of the PE criterion. This work explores the suitability of parameters $\alpha$ and $\beta$ to provide a reliable physically based classification and proposes an alternative criterion that ties all the unknowns describing meteor flight in a robust mathematical expression.

## 2. Derivation of the criterion

The PE formulation estimates the meteoroid ability to penetrate the Earth's atmosphere by taking into consideration the atmospheric density at the terminal height (the last luminous point of the meteor along its descending path through the atmosphere). Using the single body theory equations and assuming no terminal mass ($M_t$=0), an isothermal atmosphere $\rho/\rho_0$=exp(-$h/h_0$), and neglecting the average meteoroid deceleration ($<V^2>=V_e^2$), Ceplecha & McCrosky (1976) mathematically obtained:

$$-\log \rho_t = \log \sigma + \log K - 0.33 \log M_e + 2 \log V_e - \log cos(Z_R) + const.$$

*Equation 1*

where $\rho_t$ is the air density [g·cm$^{-3}$] at the meteor terminal height, $\sigma$ [s$^2$·cm$^{-2}$] is the ablation coefficient, $K$ [cm$^2$·g$^{-2/3}$] is the shape-density factor, $M_e$ [g] and $V_e$ [km·s$^{-1}$] are the entry mass and velocity of the meteoroid, respectively, and $Z_R$ [degrees] is the zenith distance of the meteoroid radiant (trajectory angle to the vertical). Note that, although for the sake of this paper we are more interested in the mathematical derivation of the PE criterion (Equation 1), Ceplecha & McCrosky (1976) initially derived the PE criterion by empirical means. Assuming that the most relevant factors regarding the ability of a meteoroid to penetrate the atmosphere are the zenith distance of the meteoroid radiant, the entry mass, and the entry velocity, Ceplecha & McCrosky (1976) grouped the impacting meteoroids implementing a least square fit on the coefficients (A, B, and C) of the following empirical expression:

$$\text{PE} = \log \rho_t + A \log M_e + B \log V_e + C \log(\cos Z_R)$$

*Equation 2*

In their work, Ceplecha & McCrosky (1976) assumed fast meteoroids, with no time to undergo significant deceleration before disintegration. This assumption led to the conclusion that most of the meteoroids in the Prairie Network data set (PN; McCrosky, Shao & Posen 1976, 1977) showed a similar ablation coefficient value (Ceplecha & McCrosky 1976). Consequently, the authors considered log($\sigma$) to be fixed and approximated its value by using the average of some subset of the analyzed PN events. As per the value of the shape variation factor ($K$) in Equation 1, its variation, as discussed by Ceplecha & McCrosky (1976), is related to the average value of log($\sigma$) of all the individual fireball observations along its trajectory. The related variation of these two parameters ($K$ and $\sigma$) were included by Ceplecha & McCrosky (1976) through a second empirical criterion which can be combined with Equation 2 to lead to the formulation of Equation 1 (for more details see Ceplecha & McCrosky 1976). Then, if log($\sigma$) is fixed for all the fireballs, it follows that log($K$) is fixed too. These two assumptions make Equation 1 and Equation 2 equivalent except for the constant (the fixed values of log($\sigma$) and log($K$)). Consequently, under these assumptions the empirical formulation (Equation 2) results in the approximate criteria illustrated in Equation 1. In any other situation, the ablation and shape variation coefficients should be explored individually for each event.

By implementing scaling laws and introducing dimensionless variables, we propose that the derivation of the adjusted coefficients of the PE criterion (Equation 1) can be replaced by a case-by-case derivation of two parameters, $\alpha$ and $\beta$, which gather all the fireball flight unknowns avoiding unnecessary assumptions. In principle, $\alpha$ depends on the pre-atmospheric cross-section-to-mass ratio (the ratio which can be easily converted into dependency on shape-density factor $K$ or bulk density, pre-atmospheric mass and shape coefficient) and trajectory slope $\gamma$ related to $Z_R$. The mass loss parameter, $\beta$, is proportional to the pre-atmospheric velocity in power of 2 and inversely proportional to the effective destruction enthalpy; it can be linked to the ablation coefficient $\sigma$ (see Gritsevich 2009). Thus, it is obvious that a similar set of physical parameters affecting the degree of deepening of meteoroids in the Earth's atmosphere is accounted for in both the mathematical formulation of the PE criterion and a combination of $\alpha$ and $\beta$ parameters. The dimensionless formulation using $\alpha$ and $\beta$ can be derived from the definition of these parameters (Gritsevich 2009):

$$\alpha = \frac{1}{2} c_d \frac{\rho_0 h_0 S_e}{M_e \sin \gamma}$$

*Equation 3*

$$\beta = (1 - \mu) \frac{c_h V_e^2}{2 c_d H^*} = (1 - \mu) \frac{V_e^2}{2} \sigma$$

*Equation 4*

Here, $c_d$ is the drag coefficient, $h_0$ [km] is a planetary scale height, $\rho_0$ [kg · m$^{-3}$] is the atmospheric density at the sea level, $S_e$ [m$^2$] is the entry middle section area of meteoroid, $M_e$ [kg] is the entry meteoroid mass, $\gamma$ [degrees] is the slope between the horizon and the meteor trajectory, $\mu$ is a shape change coefficient, $c_h$ is the heat exchange coefficient, $V_e$ [km· s$^{-1}$] is the meteoroid entry velocity, and $H^*$ [km$^2$·s$^{-2}$] is the effective destruction enthalpy.

A logarithm of the product of dimensionless parameters $\alpha$ and $\beta$ leads to:

$$\log(\alpha \cdot \beta) = \log\left(\frac{c_d \cdot \rho_0 \cdot h_0 \cdot S_e \cdot (1 - \mu) \cdot c_h \cdot V_e^2}{4 \cdot M_e \cdot \sin \gamma \cdot c_d \cdot H^*}\right) = \log\left(\frac{c_d \cdot \rho_0 \cdot h_0 \cdot S_e \cdot (1 - \mu) \cdot \sigma \cdot V_e^2}{4 \cdot M_e \cdot \sin \gamma}\right)$$

*Equation 5*

In order to enable an easier comparison between Equation 5 and the PE criterion (Equation 1), the former can be formulated with the same parametrization as the latter involving the following relationships:

- The relation: $S/S_e=(M/M_e)^\mu$, where S and M the meteoroid middle section and mass respectively (from Levin (1956, 1961)).
- The shape-factor $A=S/W^{2/3}$, where $W$ is the meteoroid volume.
- The shape-density factor $K=A \cdot c_d \cdot \rho^{-2/3}$, where $\rho$ is the meteoroid bulk density.
- The slope between the trajectory and the horizon at each time, $\gamma$, is related to $Z_R$ as $Z_R=90°-\gamma$ or $\cos(Z_R)=\sin(\gamma)$.

Including these definitions in Equation 5, it follows that:

$$\log(\alpha \cdot \beta) = \log\left(\frac{const. \cdot K \cdot \sigma \cdot (1 - \mu) \cdot V_e^2}{M_e^{(1-\mu)} \cdot \cos Z_R \cdot M^{(\mu-2/3)}}\right) =$$

$$= \log \sigma + \log K + 2 \log V_e - \log cos(Z_R) - (1 - \mu) \log M_e - \left(\mu - \frac{2}{3}\right) \log M + const.$$

*Equation 6*

While Equation 1 and Equation 6 are quite similar, there are some differences. The shape change coefficient $\mu$ is present in various terms and it is physically responsible for enhancing the mass loss along the luminous part of the trajectory that can be estimated from observations (Gritsevich & Koschny 2011). It is generally accepted that $0 \leq \mu \leq 2/3$. Nonetheless, Bouquet et al. (2014) concluded, based on their calculations using the Meteorite Observation and Recovery Project

(MORP; Halliday, Griffin & Blackwell 1996) database, that many meteoroids show $\mu=2/3$. Note, that also Ceplecha & McCrosky (1976) consider in their derivation a constant meteoroid shape that corresponds to our case of shape change coefficient $\mu=2/3$. Therefore, when $\mu=2/3$, the term in Equation 6 that depends on the mass, $M$, could be removed or considered negligible, and the $M_e$ term would be multiplied by 0.33.

Additionally, there is no terminal height related term in Equation 5. In this regard, Gritsevich & Popelenskaya (2008), Gritsevich, Lukashenko & Turchak (2016), and Moreno-Ibáñez et al. (2015, 2017) discussed the accuracy of various simplifications of the general formulation introduced by Stulov et al. (1995) to calculate the terminal height of fully ablated fireballs as well as meteorite-droppers. To match the assumptions made by Ceplecha & McCrosky (1976) (namely: no meteor deceleration, meteoroid final disintegration, constant shape, and isothermal atmosphere), the most suitable terminal height formulation simplification is the $h_I$ solution outlined in Moreno-Ibáñez et al. (2015, 2017):

$$\frac{h_I}{h_0} = -\ln\frac{\rho_t}{\rho_0} = \ln(2\alpha\beta) \Rightarrow -\log\frac{\rho_t}{\rho_0} = \log(2\alpha\beta) \Rightarrow -\log\rho_t = \log(2\alpha\beta) - \log\rho_0$$

*Equation 7*

Then, combining Equation 6 and Equation 7, it follows:

$$-\log\rho_t = \log\sigma + \log K + 2\log V_e - \log cos(Z_R) - 0.33\log M_e + const.$$

*Equation 8*

This derivation proves that Equation 1 and Equation 8, under the equal assumptions, are the same. Hence, the new approach suggested here provides a generalized physically based formulation for a robust fireball classification.

3. **Results**

These two parameters, $\alpha$ and $\beta$, are easily retrievable from the observed set of the fireball trajectory points. The values of height and velocity at these points are adjusted via the least-squares method to the scaled and dimensionless formulation of the general meteor flight equation (see Gritsevich 2009; Lyytinen & Gritsevich 2016 for further details). This methodology has been implemented for meteors showing some degree of deceleration, when at least three trajectory points (corresponding to height and velocity) are available from the observations. Given that the entry velocity is used to scale the velocity, in an ideal case, one of these three points should correspond to the entry point. However, estimating the entry velocity is generally challenging (Egal et al. 2017). In their recent study, Moreno-Ibáñez et al. (2017) discussed the viability of directly deriving the meteoroid entry velocity, along with the $\alpha$ and $\beta$ parameters.

Once the values of $\alpha$ and $\beta$ are obtained individually for each event, it is straightforward to derive the values of other meteor flight parameters. In this regard, as per Equation 4, the ablation coefficient is directly linked to the $\beta$ parameter. The value of the ablation coefficient depends on several factors, such as chemical composition, grain size, bulk density, porosity and body shape, among others, and so the generalization adopted in Ceplecha & McCrosky (1976) may induce inaccuracies in their classification. In general, the values of the ablation coefficient range between 0.001 and 0.21 $km^2 \cdot s^{-2}$ (for a review see: Silber et al. 2018), whereas for the PE criterion Ceplecha & McCrosky (1976) assumed a fixed value of $\log(\sigma)$ = -11.7 ($\sigma$ = 0.02 $km^2 \cdot s^{-2}$).

The values of $\alpha$, $\beta$, and $\sigma$ for 121 members of the PN database were computed by Gritsevich (2009). Using these data, it is possible to compare the PE results obtained from Ceplecha & McCrosky (1976) to the $\log(2\alpha\beta)$ outcome. Only for 101 of these events $\alpha$ and $\beta$, and the PE value are available, and thus these are used in the following discussion. The numerical results for both approaches are given in Table 1.

The values of the PE criterion calculated by Ceplecha & McCrosky (1976) correspond to the outcome of the empirical formulation in Equation 2. However, as discussed previously, the resulting values of Equation 2 are, in average, equivalent to the shift in the results of Equation 1. In order to avoid the effect of this shift, to set the same scale range between PE and log(2$\alpha\beta$), and to reduce the induced recurrent uncertainties due to unknown parameter approximation and measurement biases, a normalization is applied to the PE and log(2$\alpha\beta$) results. The resulting comparable histograms for both the PE criterion and the new approach are shown in Figure 1. Note that, since $h_I$ is log(2$\alpha\beta$) times a constant (Moreno-Ibáñez et al. 2015), the latter can be directly compared to PE.

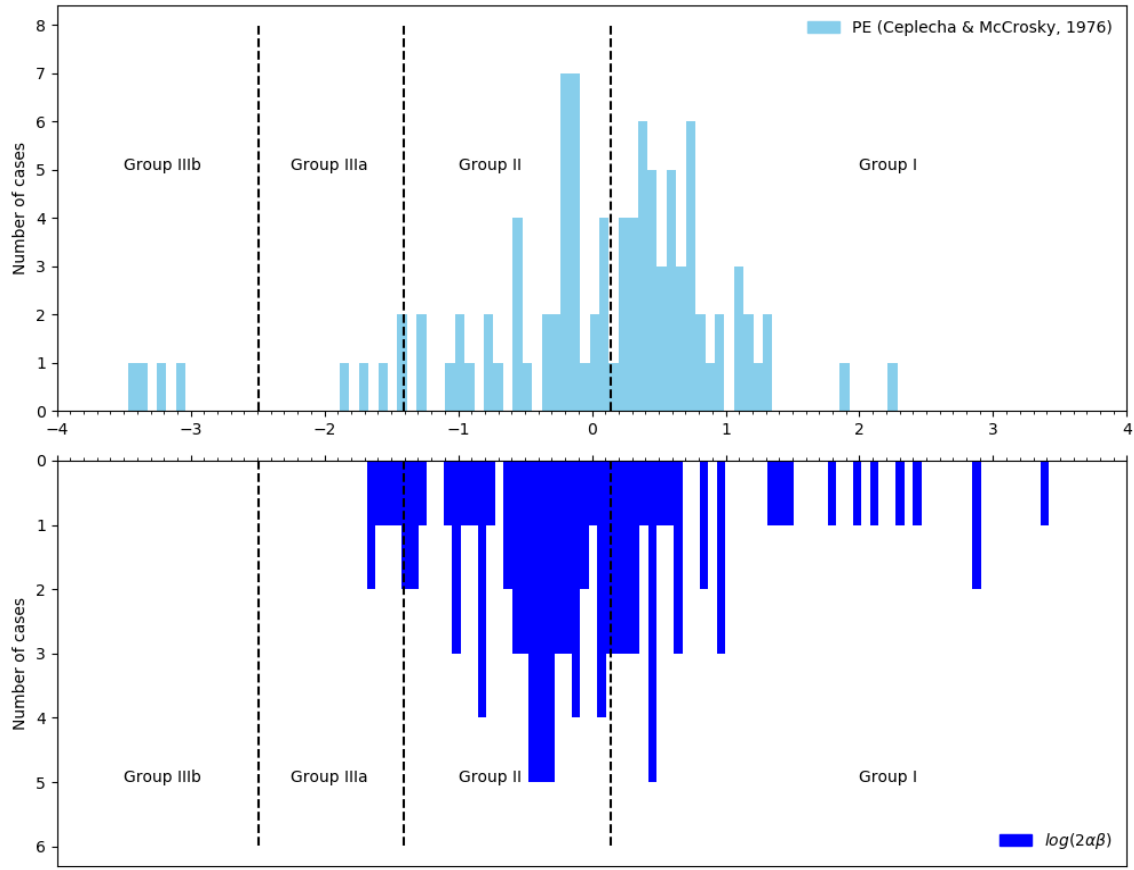

*Figure 1 – Histogram of the normalized PE (top) and log(2αβ) proposed in this study (bottom) distributions. The dashed lines delineate the four PE groups described by Ceplecha & McCrosky (1976).*

**Discussion**

From the results in Table 1, the arithmetic mean and standard deviation of the PE sample are 4.66 and $\sigma_{STDV}$=0.42, respectively. Similarly, for log(2$\alpha\beta$) the mean is 2.17 and the standard deviation approaches $\sigma_{STDV}$=0.76. The more extended distribution (larger standard deviation) obtained using the new criterion compared to the PE criterion indicates that any preset assumptions on the unknown parameters in Ceplecha & McCrosky (1976) have had a systematic notable influence on the results. Moreover, as Ceplecha & McCrosky (1976) fixed the three coefficients of their PE empirical formulation (A, B, and C, in Equation 2) according to a least square fit, the shape of the distribution is constrained by a relatively small observational dataset dictating the resulting values for these coefficients.

According to Ceplecha & McCrosky (1976), the orbital and compositional characteristics of a fireball can be classified into one of the four groups that follow the PE results for the PN data set. These groups are delineated with dashed lines in Figure 1. A subset of these data used in our study shows relevant differences in the distribution of groups I, II and IIIa. For group I the PE criterion shows 52 events, whereas the application of the log($2\alpha\beta$) criterion results in 38 events. For group II there are 40 (PE) versus 58 (log($2\alpha\beta$)) occurrences. For group IIIa it results in 4 (PE) against 5 (log($2\alpha\beta$)). Interestingly, the log($2\alpha\beta$) criterion does not show any event in group IIIb.

The 4 events present in group IIIb using the PE criterion illustrate how the observational assumptions required to use this criterion could effectively affect the grouping of fireballs. An example of this are two of these four events (events 39043.601 and 39048.956 in Table 1) which belong to the Draconids meteor shower. Their entry velocities (23.7 km/s for fireball 39043.601, and 22.1 km/s for fireball 39048.956) are within the historically observed values for these meteors (~21 km/s; see Trigo-Rodríguez et al. 2013). These meteoroids are expected to be very fragile and so it seems reasonable their classification within the group IIIb. However, their calculated terminal heights, $h_l$, indicate that these bodies reached anomalous low heights (57.9 km for fireball 39043.601, and 66.8 km for fireball 39048.956) compared to the general behaviour of this meteor shower (85 - 90 km; Trigo-Rodríguez et al. 2013). While the accuracy of the calculated terminal height depends on the observed mass loss (the $\beta$ values; see Moreno-Ibáñez et al. 2015), the $\beta$ values derived for the event 39043.601 ($\beta = 3.457$) and for the event 39048.956 ($\beta = 30.453$) ensure a low discrepancy between the terminal height observed and the one calculated. Thus, if the observations are reliable the terminal heights calculated are reliable too, and these events do not longer belong to group IIIb as it is defined by Ceplecha & McCrosky (1976). The new accuracy level obtained with the new formulation of the PE criterion requests a new classification scheme to better understand the event under study. Furthermore, the case of these Draconids meteors is also an example of how our new criterion is more consistent, and less influenced by data biased interpretation.

The shift in the group distribution of the data set according to the new criterion, along with the larger standard deviation of log($2\alpha\beta$) are indicative of an induced shift towards group II of the original PE fireball classification. This may indicate a larger number of carbonaceous chondrites within the Prairie Network data set than what is showed by the log($2\alpha\beta$) alternative. This result contradicts one of the main goals of Ceplecha & McCrosky (1976) research, as they tried to set a criterion to clearly identify ordinary chondrites (represented by group I) through fireball observations. Further analysis on this regard should include a revision on the limits between fireball groups, but this is out of the scope of this paper and we propose it as a subject for future investigations.

Although both Equation 1 and Equation 6 are derived from the basic meteor physics equations, their main difference emerges from the way of using them. While the input to Equation 1 are parameters retrieved from meteor observations and empirically set assumptions, the use of scaling laws and dimensionless variables requires the derivation of two parameters (Equation 6) based only on the observables. It is quite remarkable though that, as discussed by Moreno-Ibáñez et al. (2015), the simplified solution $h_l$ (Equation 7) does not provide the best adjustment to the observed terminal heights. Hence, as Equation 1 and Equation 8 are equivalent, the PE criterion may suffer from the same level of inaccuracies. By investigating mainly meteorite- and impact crater-producing criteria, Gritsevich et al. (2012) initially suggested that a combination of $\alpha$ and $\beta$ could outline an alternative fireball classification, removing unknown meteor flight parameter assumptions from the criterion. Moreno-Ibáñez et al. (2015, 2017) also suggested that the derivation of other meteor flight parameters (i.e., terminal height) using the new proposed methodology could identify possible empirical inaccuracies, and therefore identify outliers in large data sets. In these cases, it could be convenient to correlate suitable parameters with $\alpha$ or $\beta$ to explore these inaccuracies depending on the data set or event under study. Here we have demonstrated that the use of these parameters in Equation 6 provide a more reliable generalized criterion for fireball classification and does not require prior knowledge about meteor parameters or empirical coefficients.

The PE criterion outlined by Ceplecha & McCrosky (1976) is commonly used as an informative tool to get a quick overview of the event under study. Further study of the event requires the implementation of dynamical and/or photometric approaches to derive more properties of the phenomena. In many cases these approaches forcedly need the assumption of some variables. Indeed, the derivation of the entry mass ($M_e$ in Equation 2) may lead to different results when a dynamical or photometric approach are implemented for its derivation (Gritsevich 2008c). In addition to this, the impact of meteoroids to the atmosphere involves more complex phenomena that hinder the analytical study (see e.g. Öpik 1958; Bronshten 1983). For instance, large meteoroids could experience severe fragmentation that may split the initial single body into several bodies thus affecting their atmospheric trajectories and brightness. For these cases, a more detailed and cautious study is needed, and the classification using the PE criterion, when viable, is more challenging.

The new formulation presented in this paper provides reliable fireball classification which can, in some cases, be adapted for complex fireball scenarios. The relative ease to obtain the values of $\alpha$ and $\beta$ from observations have been demonstrated by recent studies. Indeed, Gritsevich (2009) derived the $\alpha$ and $\beta$ parameters for 143 fireball events from the Meteor Observation and Recovery Project (MORP, Canada; Halliday et al.1996) and 121 events from the Prairie Network database (PN, USA; McCrosky et al. 1976, 1977). Additionally, a set of the 278 fireball events observed by the Desert Fireball Network (DFN, Australia; Howie et al., 2017) was analyzed in Sansom et al. (2019) to show how the visualization in an $\alpha$–$\beta$ diagram can quickly identify which fireballs may produce meteorites. These studies allowed the first and an efficient large-scale classification of meteors, the identification of meteorite-producing fireballs, and the capacity to forecast impact-crater production using accurate parameters. The graphical comparison of meteor events in the ln $\alpha$ - ln $\beta$ plane reveals that meteorite-droppers lie in a different region than fully ablated fireballs (Gritsevich et al. 2012; Sansom et al. 2019). This conclusion is even more pronounced if the trajectory slope is taken into account in a comparative plot (Sansom et al. 2019). For instance, Figure 2, which is an updated version of the Gritsevich et al. (2012) diagram, shows how meteorite-producing events (such as Innisfree, Lost City, Annama, Bunburra Rockhole, etc.) cluster in a defined localized region of the diagram compared to the rest of the fireballs. Moreover, the location of the few identified carbonaceous chondrites of the MORP database (northern and southern Taurids) fall in another delimited area of the diagram.

The lack of reliability of the PE criterion is clearly evidenced when the ablation coefficient is evaluated. In the scaling and dimensionless approach this value is obtained from observations and treated as a free variable. As it can be derived from the data set in Table 1, the fixed value used in the original PE criterion ($\sigma = 0.02$ km$^2 \cdot$s$^{-2}$) is far from the averaged value obtained here directly from observations ($\sigma = 0.013$ km$^2 \cdot$s$^{-2}$). This issue is partially solved by Ceplecha & McCrosky (1976) introducing a second criterion that accounts for the average variations in log($\sigma$) and log($K$). It should be noted that this second criterion is rarely used in the literature.

Furthermore, the single body theory includes an exponential atmosphere to derive the meteor flight equations. To support the analytical derivation, this simplification is also implemented in the scaling laws and dimensionless variables formulation. However, as discussed by Lyytinen & Gritsevich (2016), this latter formulation is flexible enough to allow the incorporation of any other atmospheric model (e.g., MSIS-E 90 (Hedin 1991) or actual weather station information at the time and location of the fireball) to adjust the atmospheric pressure to actual values and enhance the accuracy of the outcome. Moreover, it does not require any formulation adaptation, which ultimately permits re-assigning the classification of individual events or complete data sets with little effort by utilizing a desired atmospheric model.

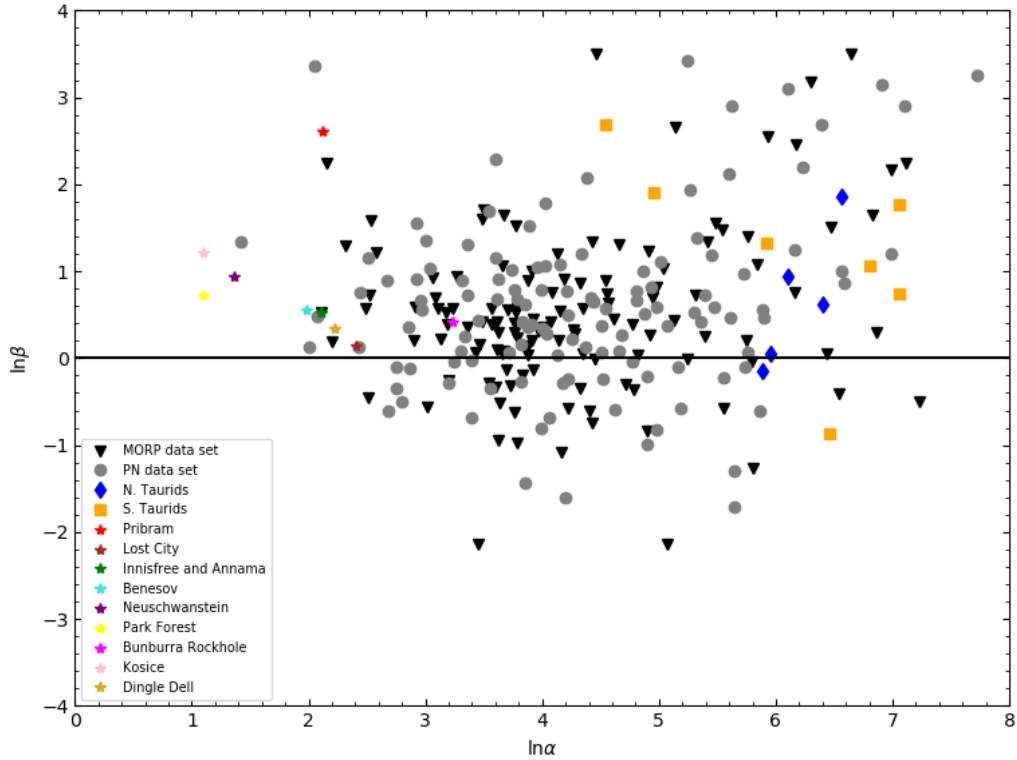

*Figure 2 - Diagram showing the lnα and lnβ for MORP and PN database events based on Gritsevich et al. (2012), where the few Taurids registered by the MORP are marked separately. Symbols for Pribram, Lost City, Innisfree, Neuschwanstein (Gritsevich 2008a), Benesov (Gritsevich 2008b; Gritsevich et al. 2017) Park Forest (Meier et al. 2017), Annama (Trigo-Rodríguez et al. 2015), Bunburra Rockhole (Sansom et al. 2015), Kosice (Gritsevich et al. 2017), and Dingle Dell (Devillepoix et al. 2018) meteorites are also shown. Note that Innisfree and Annama show nearly the same α and β values and so their symbols coincide in the graph.*

Up to this point we have accepted $\mu$ (Levin 1956, 1961) as fixed with a value of 2/3. Although this is valid for most cases, an accurate analysis of this parameter is necessary in some studies. The parameters $\alpha$ and $\beta$ are retrieved directly from the observations, but individual meteor flight parameters are derived from these two values. For instance, the ablation coefficient, $\sigma$, is obtained from the $\beta$ parameter and, as per Equation 4, the derivation is dependant of the $\mu$ value. Understanding the value of $\mu$ at each event is essential to use a more general formulation (Equation 6) and achieve a greater accuracy and detail in the classification.

Finally, the values of $\alpha$ or $\beta$ can be also retrieved accurately under certain meteoroid atmospheric fragmentation scenarios (see Gritsevich 2008a; Moreno-Ibáñez 2018). It has been demonstrated that these parameters can be used to identify meteorite-producing fireballs as well as to predict formation of impact craters for larger impactors (Gritsevich et al. 2012; Turchak & Gritsevich 2014; Sansom et al. 2019). On one hand, this allows for the new criterion to be used for meteorite-droppers (and not only to disintegrating fireballs); and on the other hand, it provides a reliable classification for the atmospheric behaviour of meter-sized projectiles.

## 4. Conclusions

We have revised the PE criterion, introduced by Ceplecha & McCrosky (1976), that has been extensively used to interpret the nature and origin of impacting meteoroids. The grouping of fireballs based on this criterion relies on trajectory analysis followed by an elemental approach and simplifications embodied in the single body theory. This empirical formulation of the criterion along with the introduction of fixed constants for meteors showing little to no deceleration, and the application of the criterion to just one data set acknowledges statistical bias on the results of Ceplecha & McCrosky (1976). This work discussed an alternative approach. The main conclusions found in this study can be summarized as follows:

1) The alternative formulation described here is based on the scaling laws and dimensionless variables. It offers a generalized formulation of the PE criterion (Equation 6) avoiding the original PE assumptions on the meteor flight unknowns and making it unnecessary to introduce empirical constants.
2) Our results suggest that the application of the PE criterion in its original form (in part due to pre-set ablation coefficient value) causes an artificial shift in the PE classification towards group II proposed by Ceplecha & McCrosky (1976). The reduction of inaccuracies in our new formulation for the PE criterion shall lead to the use of an alternative classification scheme and grouping. The PE criterion was envisioned as an informative tool and so shall the original classification groups be considered.
3) The approach suggested in this work allows for a broader, reliable and efficient classification in the entire range of meteor events from fast cometary stream particles to slow and durable meteorite-droppers that are capable of penetrating deep into the atmosphere, and as such can be source of hazard to humans.

As future work, the new $\log(2\alpha\beta)$ criterion should be applied to additional meteorite falls to compare the bulk physical properties (density, strength, etc.) of the recovered meteorites with the expected nature of the incoming meteoroids.

**Acknowledgments**

MMI acknowledges the support of Aistech Space S.L. MG acknowledges the Academy of Finland project no. 325806 (PlanetS). Research at the Ural Federal University is supported by the Russian Foundation for Basic Research, project nos. 18-08-00074 and 19-05-00028 and the Act 211 of the Government of the Russian Federation, agreement No 02.A03.21.0006. JMTR acknowledges the support of MEC under AYA research grant PGC2018-097374-B-I00. The authors sincerely thank the reviewers for their valuable feedback which helped us to improve the manuscript.

**References**

Bouquet A., Baratoux D., Vaubaillon J., Gritsevich M., Mimoun D., Mousis O., Bouley S., 2014, Planet. Space Sci., 103, 238

Bronshten V.A., 1983, Physics of Meteoric Phenomena, Kluwer, Dordrecht, Holland, 356 pp.

Brown P., Marchenko V., Moser D. E., Weryk R., Cooke W., 2013, Meteorit. Planet. Sci., 48, 270

Ceplecha Z., McCrosky R. E., 1976, JGR, 81, 6257

Devillepoix H.A.R. et al., 2018, Meteorit. Planet. Sci., 53 (10), 2212

Dmitriev V., Lupovka V., Gritsevich M., 2015, Planet. Space Sci., 117, 223

Egal A., Gural P. S., Vaubaillon J., Colas F., Thuillot, W., 2017, Icarus, 294, 43

Gritsevich M. I., 2007, Sol. Syst. Res., 41, 509

Gritsevich M.I., 2008a, Sol. Syst. Res., 42(5), 372

Gritsevich M.I., 2008b, Moscow Uni. Mech. Bull., 63(1), 1

Gritsevich M.I., 2008c, Doklady Physics, 53(2), 97

Gritsevich M.I., 2009, Adv. Space Res., 44, 323

Gritsevich M., Koschny D., 2011, Icarus, 212(2), 877

Gritsevich M.I., Popelenskaya N. V., 2008, Doklady Physics, 53, 88

Gritsevich M.I., Stulov V.P., 2006, Sol. Syst. Res., 40(6), 477

Gritsevich M.I., Stulov V.P., Turchak L.I., 2009, Doklady Physics, 54(11), 499

Gritsevich M. I., Stulov V. P., Turchak L. I., 2012, Cosmic Res., 50, 56

Gritsevich M.I., Lukashenko V.T., Turchak L.I., 2016, Math. Models Comput. Simul., 8(1), 1

Gritsevich M. et al., 2017, in Trigo-Rodríguez J. M., Gritsevich M., & Palme J., eds., Assessment and Mitigation of Asteroid Impact Hazards: Proceedings of the 2015 Barcelona Asteroid Day, Springer International Publishing Switzerland, Vol. 46, p. 153.

Halliday, I., Griffin A. A., Blackwell A. T.,1996, Meteorit. Planet. Sci., 31, 185

Hedin A. E. 1991, JGR, 96, 1159

Howie R. M., Paxman J., Bland P. A., Towner M. C., Cupak M., Sansom E. K., Devillepoix H. A., 2017, Exp. Astron., 1

Levin B. J., 1956, Bull. Astron. Inst. Czech., 7, 58

Levin B.I., 1961, Physikalische Theorie der Meteore und die meteoritische Substanz im Sonnensystem, Berlin, Akademie-Verlag

Lyytinen E., Gritsevich M., 2016, Planet. Space Sci., 120, 35

Matlovič P., Tóth J., Rudawska R., Kornoš L., 2017, Planet. Space Sci., 143, 104

McCrosky R.E., Shao C.-Y., Posen A., 1976, No. 665, Center for Astrophysics, Cambridge, Massachusetts

McCrosky R.E., Shao C.-Y., Posen A.,1977, Preprint No. 721, Center for Astrophysics, Cambridge, Massachusetts

Meier M.M.M., Welten K.C., Riebe M.E.I., Caffee M.W., Gritsevich M., Maden C., Busemann H., 2017, Meteorit. Planet. Sci., 52, 1561

Moreno-Ibáñez M., 2018, PhD Thesis, Universitat Autònoma de Barcelona

Moreno-Ibáñez M., Gritsevich M., Trigo-Rodríguez J. M., 2015, Icarus, 250, 544

Moreno-Ibáñez M., Gritsevich M., Trigo-Rodríguez J. M., 2016, in Roggemans A., Roggemans P., eds., Proceedings of the International Meteor Conference. International Meteor Organization, Egmond, the Netherlands, p. 192

Moreno-Ibáñez M., Gritsevich M., Trigo-Rodríguez J. M. 2017, in Trigo-Rodríguez J. M., Gritsevich M., & Palme J., eds., Assessment and Mitigation of Asteroid Impact Hazards: Proceedings of the 2015 Barcelona Asteroid Day, Springer International Publishing Switzerland, Vol. 46, p. 129

Öpik E.J., 1958, Physics of Meteor Flight in the Atmosphere (New York: Interscience), 1–179

Sansom E. K., Bland P. A., Paxman J., Towner M. C., 2015, Meteorit. Planet. Sci., 50(8), 1423

Sansom E.K. et al., 2019, ApJ, 885 (2): 115

Silber E. A., Boslough M., Hocking W. K., Gritsevich M., Whitaker R.W., 2018, Adv. Space Res., 62(3), 489

Stulov V. P., 1997, Applied Mech. Rev., 50, 671

Stulov V.P., Mirskii A.N., Vislyi A.I., 1995, Aerodinamika bolidov (Aerodynamics of Bolides) (Moscow: Nauka)

Trigo-Rodríguez J.M., 2019, in Colonna G., Capitelli M., Laricchiuta A., eds., Hypersonic Meteoroid Entry Physics, Institute of Physics Publishing, IOP Series in Plasma Physics, p. 4-1

Trigo-Rodríguez J.M., Llorca, J., 2006. MNRAS, 372, 655

Trigo-Rodríguez J.M., Llorca, J., 2007. MNRAS, 375, 415

Trigo-Rodríguez J.M. et al., 2013, MNRAS 433, 560

Trigo-Rodríguez J.M. et al., 2015, MNRAS, 449 (2), 2119

Turchak L.I., Gritsevich M.I., 2014, J. Theor. Appl. Mech., 44(4), 15

Vaubaillon J. et al., 2015, Earth Moon Planets, 114(3-4), 137

# TABLES

| ID | α | β | σ (km²·s⁻²) | log(2αβ) | $h_l$ (km) | PE |
|---|---|---|---|---|---|---|
| **38737.703** | 18.59 | 2.478 | 0.017 | 1.964 | 32.3 | -4.8 |
| **38768.668** | 12.3 | 3.194 | 0.021 | 1.895 | 31.2 | -4.81 |
| **38850.755** | 122.48 | 2.180 | 0.007 | 2.728 | 44.9 | -4.62 |
| **39031.636** | 48.06 | 1.421 | 0.007 | 2.135 | 35.2 | -4.26 |
| **39038.719B** | 41.83 | 2.780 | 0.020 | 2.367 | 39.0 | -5.05 |
| **39043.601** | 473.24 | 3.457 | 0.012 | 3.515 | 57.9 | -6.12 |
| **39048.956** | 188.37 | 30.453 | 0.125 | 4.060 | 66.8 | -5.96 |
| **39055.568** | 43.08 | 2.200 | 0.017 | 2.278 | 37.5 | -4.4 |
| **39065.674** | 37.25 | 1.965 | 0.013 | 2.166 | 35.7 | -4.62 |
| **39078.796** | 83.24 | 1.991 | 0.034 | 2.520 | 41.5 | -4.72 |
| **39080.635A** | 47.15 | 1.851 | 0.004 | 2.242 | 36.9 | -4.37 |
| **39093.754B** | 63.94 | 2.923 | 0.017 | 2.573 | 42.4 | -5.33 |
| **39113.752A** | 11.43 | 2.125 | 0.019 | 1.686 | 27.8 | -4.71 |
| **39113.586B** | 150.36 | 3.013 | 0.011 | 2.957 | 48.7 | -4.36 |
| **39115.957A** | 134.16 | 0.813 | 0.003 | 2.339 | 38.5 | -4.35 |
| **39120.796** | 78.63 | 1.129 | 0.004 | 2.249 | 37.0 | -4.37 |
| **39122.638** | 68.07 | 0.787 | 0.003 | 2.030 | 33.4 | -4.87 |
| **39128.994** | 108.13 | 1.311 | 0.015 | 2.453 | 40.4 | -4.5 |
| **39129.7714** | 61.78 | 1.033 | 0.013 | 2.106 | 34.7 | -4.61 |
| **39130.705** | 53.99 | 2.186 | 0.006 | 2.373 | 39.1 | -4.45 |
| **39138.854A** | 7.74 | 28.993 | 0.093 | 2.652 | 43.7 | -4.74 |
| **39138.996B** | 20.76 | 2.807 | 0.006 | 2.066 | 34.0 | -5.26 |
| **39143.902** | 34.72 | 5.395 | 0.026 | 2.574 | 42.4 | -4.71 |
| **39154.662** | 19.16 | 1.957 | 0.009 | 1.875 | 30.9 | -4.36 |
| **39169.852** | 56.4 | 1.323 | 0.005 | 2.174 | 35.8 | -4.49 |
| **39182.832** | 19.38 | 1.736 | 0.011 | 1.828 | 30.1 | -4.29 |
| **39197.813** | 220.75 | 2.067 | 0.005 | 2.960 | 48.7 | -4.54 |
| **39224.744** | 93.97 | 1.776 | 0.003 | 2.523 | 41.5 | -4.14 |
| **39240.872** | 15.63 | 0.705 | 0.005 | 1.343 | 22.1 | -4.31 |
| **39265.77** | 76.8 | 3.348 | 0.006 | 2.711 | 44.6 | -4.53 |
| **39276.818** | 597.96 | 14.599 | 0.044 | 4.242 | 69.8 | -6.08 |
| **39317.812** | 36.54 | 9.820 | 0.057 | 2.856 | 47.0 | -5.45 |
| **39360.701** | 449.37 | 22.270 | 0.078 | 4.301 | 70.8 | -4.79 |
| **39363.818** | 37.42 | 2.492 | 0.007 | 2.271 | 37.4 | -4.4 |
| **39368.873** | 28.77 | 3.674 | 0.017 | 2.325 | 38.3 | -4.7 |
| **39382.599** | 105.84 | 1.080 | 0.002 | 2.359 | 38.8 | -4.37 |
| **39404.832** | 27.98 | 1.292 | 0.011 | 1.859 | 30.6 | -4.65 |
| **39405.62** | 131.67 | 1.675 | 0.017 | 2.645 | 43.5 | -4.89 |
| **39406.784A** | 14.42 | 2.448 | 0.017 | 1.849 | 30.4 | -5.1 |
| **39409.825** | 121.71 | 1.936 | 0.004 | 2.673 | 44.0 | -4.17 |
| **39411.697A** | 361.63 | 1.749 | 0.003 | 3.102 | 51.1 | -4.49 |
| **39423.723** | 212.33 | 1.525 | 0.003 | 2.811 | 46.3 | -4.72 |
| **39434.946** | 17.43 | 1.440 | 0.014 | 1.701 | 28.0 | -5.19 |

| ID | α | β | σ (km²·s⁻²) | log(2αβ) | $h_l$ (km) | PE |
|---|---|---|---|---|---|---|
| 39439.905A | 310.92 | 0.907 | 0.001 | 2.751 | 45.3 | -4.36 |
| 39460.599A | 307.28 | 2.633 | 0.006 | 3.209 | 52.8 | -4.55 |
| 39469.850A | 727.62 | 2.388 | 0.007 | 3.541 | 58.3 | -5.37 |
| 39476.67 | 43.8 | 1.988 | 0.010 | 2.241 | 36.9 | -4.48 |
| 39478.824A | 233.96 | 3.275 | 0.007 | 3.185 | 52.4 | -4.42 |
| 39494.62 | 200.98 | 1.684 | 0.010 | 2.831 | 46.6 | -4.91 |
| 39499.829 | 54.85 | 1.416 | 0.018 | 2.191 | 36.1 | -4.62 |
| 39512.906 | 17.6 | 0.885 | 0.004 | 1.493 | 24.6 | -4.58 |
| 39533.668 | 66.26 | 0.202 | 0.001 | 1.428 | 23.5 | -4.18 |
| 39534.814 | 259.44 | 0.805 | 0.001 | 2.621 | 43.1 | -4.47 |
| 39535.979E | 48.63 | 4.585 | 0.005 | 2.649 | 43.6 | -4.11 |
| 39538.928B | 1220.96 | 18.087 | 0.008 | 4.645 | 76.5 | -5.0 |
| 39559.976A | 90.71 | 1.451 | 0.007 | 2.420 | 39.8 | -4.2 |
| 39608.773 | 178.07 | 0.558 | 0.003 | 2.298 | 37.8 | -4.74 |
| 39667.805 | 159.13 | 1.463 | 0.016 | 2.668 | 43.9 | -4.74 |
| 39681.791 | 20.28 | 3.861 | 0.018 | 2.195 | 36.1 | -5.21 |
| 3939728.5 | 80.43 | 7.989 | 0.057 | 3.109 | 51.2 | -4.98 |
| 39730.842 | 145.28 | 0.443 | 0.005 | 2.110 | 34.7 | -4.9 |
| 39755.621A | 85.1 | 1.920 | 0.020 | 2.514 | 41.4 | -6.01 |
| 39794.700B | 18.7 | 4.752 | 0.020 | 2.250 | 37.0 | -4.75 |
| 39812.752 | 45.74 | 0.762 | 0.007 | 1.843 | 30.3 | -4.51 |
| 39815.653 | 29.93 | 0.977 | 0.011 | 1.767 | 29.1 | -4.51 |
| 39820.556A | 83.58 | 1.956 | 0.006 | 2.515 | 41.4 | -4.27 |
| 39827.608A | 91.55 | 0.786 | 0.007 | 2.158 | 35.5 | -4.7 |
| 39833.53 | 1079.86 | 3.324 | 0.007 | 3.856 | 63.5 | -5.07 |
| 39850.552 | 54.21 | 0.444 | 0.004 | 1.682 | 27.7 | -4.36 |
| 39856.722 | 70.84 | 1.247 | 0.002 | 2.247 | 37.0 | -3.7 |
| 39860.773C | 46.1 | 1.514 | 0.015 | 2.145 | 35.3 | -4.73 |
| 39863.941B | 204.2 | 3.994 | 0.005 | 3.212 | 52.9 | -4.72 |
| 39874.586 | 49.76 | 1.482 | 0.007 | 2.169 | 35.7 | -4.74 |
| 39888.872 | 2256.58 | 25.797 | 0.010 | 5.066 | 83.4 | -4.91 |
| 39907.858 | 507.97 | 8.988 | 0.005 | 3.961 | 65.2 | -4.95 |
| 39921.766C | 16.31 | 0.606 | 0.006 | 1.296 | 21.3 | -4.77 |
| 39935.621 | 14.61 | 0.543 | 0.003 | 1.200 | 19.8 | -4.32 |
| 39957.75 | 175.42 | 0.897 | 0.002 | 2.498 | 41.1 | -4.4 |
| 39972.692 | 281.86 | 0.272 | 0.002 | 2.186 | 36.0 | -4.46 |
| 39985.801A | 316.71 | 1.062 | 0.005 | 2.828 | 46.6 | -4.55 |
| 40040.653 | 1002.52 | 23.310 | 0.017 | 4.670 | 76.9 | -5.25 |
| 40065.782B | 272.67 | 1.602 | 0.007 | 2.941 | 48.4 | -4.74 |
| 40120.747A | 118.74 | 0.965 | 0.004 | 2.360 | 38.9 | -4.57 |
| 40151.609 | 15.59 | 0.902 | 0.010 | 1.449 | 23.9 | -4.66 |
| 40161.736 | 27.57 | 2.459 | 0.017 | 2.132 | 35.1 | -4.42 |
| 40261.914 | 45.54 | 1.179 | 0.007 | 2.031 | 33.4 | -4.19 |
| 40298.826B | 139.48 | 2.268 | 0.005 | 2.801 | 46.1 | -4.35 |
| 40330.789A | 353.85 | 0.548 | 0.001 | 2.589 | 42.6 | -4.2 |

| ID | α | β | σ (km²·s⁻²) | log(2αβ) | $h_l$ (km) | PE |
|---|---|---|---|---|---|---|
| **40353.738B** | 284.02 | 0.180 | 0.001 | 2.010 | 33.1 | -4.46 |
| **40403.719** | 145.38 | 1.799 | 0.006 | 2.719 | 44.8 | -4.67 |
| **40405.773** | 28.94 | 2.071 | 0.019 | 2.079 | 34.2 | -4.77 |
| **40456.735** | 710.42 | 2.740 | 0.005 | 3.590 | 59.1 | -4.5 |
| **40503.803** | 7.44 | 1.142 | 0.005 | 1.230 | 20.3 | -5.08 |
| **40590.593** | 11.34 | 1.129 | 0.011 | 1.408 | 23.2 | -4.44 |
| **40617.888** | 25.8 | 0.957 | 0.011 | 1.694 | 27.9 | -4.56 |
| **40660.676B** | 31.64 | 1.542 | 0.005 | 1.989 | 32.8 | -3.87 |
| **40755.77** | 56.19 | 5.986 | 0.044 | 2.828 | 46.6 | -4.45 |
| **40806.72** | 47.08 | 0.240 | 0.003 | 1.354 | 22.3 | -4.54 |
| **40934.927** | 36.75 | 3.174 | 0.032 | 2.368 | 39.0 | -4.53 |
| **40996.84** | 29.85 | 0.507 | 0.005 | 1.481 | 24.4 | -4.12 |
| **41014.777** | 237.49 | 1.804 | 0.012 | 2.933 | 48.3 | -4.48 |

*Table 1 – α, β, and σ parameters (Gritsevich 2009), log(2αβ), terminal height $h_l$ (Moreno-Ibáñez et al. 2015) and PE (Ceplecha & McCrosky 1976) values for 101 fireballs of the Prairie Network. The ID follows the notation introduced by Ceplecha & McCrosky (1976).*